\documentclass{iaus}
\usepackage{graphicx}

\title[Upper limits on the mass of SMBHs] %% give here short title %%
{Upper limits on the mass of supermassive black holes from
  STIS archival data}

\author[Beifiori et al.]  %% give here short author list %%
{A. Beifiori$^1$, E.~M. Corsini$^1$, E. Dalla Bont\`a$^1$,
 A. Pizzella$^1$, \break 
 L. Coccato$^2$, M. Sarzi$^3$, \and F. Bertola$^1$}

\affiliation{$^1$Dipartimento di Astronomia, Universit\`a di Padova, vicolo dell'Osservatorio 3, I-35122 Padova, Italy 
\break email: alessandra.beifiori@unipd.it,
enricomaria.corsini@unipd.it, elena.dallabonta@unipd.it, alessandro.pizzella@unipd.it, francesco.bertola@unipd.it\\[\affilskip]
$^2$Max-Planck-Institut f\"{u}r extraterrestrische Physik, Giessenbachstrasse 1, D-85748 Garching bei M\"{u}nchen, Germany \break
email: lcoccato@mpe.mpg.de \\[\affilskip]
$^3$Centre for Astrophysics Research, University of Hertfordshire, College Lane, Hatfield AL10 9AB, UK \break email: m.sarzi@herts.ac.uk
}

\pubyear{2007}
\volume{245}  %% insert here IAU Symposium No.
\pagerange{119--126}
\date{?? and in revised form ??}
\jname{Formation and Evolution of Galaxy Bulges}
\editors{M. Bureau, E. Athanassoula \& B. Barbuy, eds.}

\begin{document}

\def\kms{$\mathrm{km\;s}^{-1}$}
\def\dg{$^\circ$}
\def\ha{H$\alpha$}
\def\ss{$\sigma_c$}
\def\msun{M$_{\odot}$}
\def\ms{$M_\bullet-\sigma_c$}
\def\mbh{$M_\bullet$}
\def\nii{[N~{\small II}]}
\def\niipg{[N~{\small II}]$\,\lambda\lambda6548,6583$}
\def\sii{[S~{\small II}]}
\def\siipg{[S~{\small II}]$\,\lambda\lambda6716,6731$}
\def\farcs{\hbox{$.\!\!^{\prime\prime}$}}
\def\farcsn{\hbox{$\ \!\!^{\prime\prime}$}}

\maketitle

\begin{abstract}
The growth of supermassive black holes (SMBHs) appears
to be closely linked with the formation of spheroids. There is a
pressing need to acquire better statistics on SMBH masses, since the
existing samples are preferentially weighted toward early-type
galaxies with very massive SMBHs. With this motivation we started a
project aimed at measuring upper limits on the mass of the SMBHs that
can be present in the
center of all the nearby galaxies ($D<100$ Mpc) for which STIS/G750M
spectra are available in the HST archive. These upper limits will be
derived by modeling the central emission-line widths (\niipg, \ha\
and \siipg) observed over an aperture of $\sim$0\farcs1 ($R<50$ pc). Here we present our
preliminary results for a subsample of 76 bulges.

\keywords{black hole physics, galaxies: kinematics and dynamics, 
galaxies: structure}
\end{abstract}

\firstsection % if your document starts with a section,

\section{Introduction}

The census of supermassive black holes (SMBHs) is large enough to probe
the links between mass of SMBHs and the global properties of the host
galaxies (see for a review \cite[Ferrarese \& Ford 2005]{FerFord2005}).
However, accurate measurements
of masses \mbh\ of SMBHs are available for a few tens of galaxies
and the addition of new determinations is highly desirable.
To this purpose we started a project aimed at measuring upper limits
on \mbh\ of SMBHs that can be present in the center of all the nearby galaxies ($D<100$ Mpc)
for which STIS/G750M spectra are available in the HST archive. 
We retrieved data for 182 galaxies spanning over all the morphological
types. This will extend previous works by \cite{Sarzi2002} and
\cite{Verdoes2006}. 
Here, we show the preliminary results for a subsample of 76 bulges
with stellar velocity dispersion \ss\ available in literature.

\firstsection

\section{Data reduction and analysis}

\begin{figure}[]
\begin{center}
\includegraphics[height=6cm]{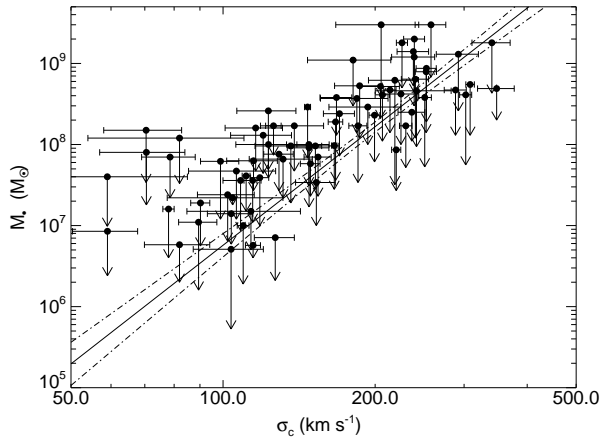}
\end{center}
\caption{Comparison between our \mbh\ upper limits (filled
   circles) and the \ms\ relation by \cite{FerFord2005}. The edges of the
   arrows correspond to upper limits obtained with $i=33$\dg\ and
   $i=81$\dg, respectively. 
}
\label{ms}
\end{figure}

The STIS/G750M spectra were obtained with either the 0\farcs2 $\times$
52\farcsn\ or 0\farcs1 $\times$ 52\farcsn slit crossing the galaxy nucleus along a random
position angle. 
The observed spectral region includes the \niipg, \ha, \siipg\ emission lines.
Routine data reduction was performed. For each galaxy we
obtained the nuclear spectrum by extracting a 0\farcs25-wide aperture
centered on the continuum peak. 
The ionized-gas velocity dispersion was measured by fitting
Gaussians to the broad and narrow components of the observed lines.
We assumed that the ionized gas resides in a thin disk and moves into
circular orbits. The local circular velocity is dictated by the
gravitational influence of the putative SMBH.  In our model we
neglected non-gravitational forces as well as the mass contribution of
the stellar component.  By taking into account for these effects, we
would obtain tighter upper limits on \mbh.  To derive the upper
limits on \mbh\ we first built the gaseous velocity field and
projected it onto the sky plane according to the disk
orientation. Then we observed it by simulating the actual setup of
STIS as done by \cite{Sarzi2002}.  We have no information on the
orientation of the gaseous disk within the central
aperture. Therefore, to reproduce the observed line width and flux radial
profile we adopted two different inclinations of the
gaseous disk that correspond to 68\% upper and lower confidence limits
for the \mbh\ for randomly orientated disks (Figure \ref{ms}).

\firstsection

\section{Results}
For some galaxies of the sample, either upper limits
(\cite[Sarzi et al. 2002]{Sarzi2002}; \cite[Coccato et al. 2006]{Coccato2006}) or accurate measurements of \mbh\
(\cite[Ferrarese \& Ford 2005]{FerFord2005}) were available. They are consistent with 
our measurements in that our \mbh\ upper limits always 
exceed the corresponding \mbh\ determinations.
Therefore, we are confident to obtain reliable estimates
of the upper limit on \mbh\ for all the remaining galaxies of our sample.
This method will allow to increase the statistical significance of the
relationships between \mbh\ and galaxy properties. Adding new masses
in both the upper and lower ends of \ms\ will allow to identify
peculiar objects worthy of further investigations. 
The resulting upper limits are close to the \ms\ by \cite{FerFord2005}.
At small \ss\ most of \mbh\ are above \ms. This can be explained in
term of non-gravitational forces (\cite[Sarzi et al. 2002]{Sarzi2002}). At larger \ss\
some of \mbh\ fall below the \ms\ relation. They could be the laggard
SMBHs discussed by \cite{Vittorini2005}.

\firstsection

\end{document}